\documentclass[letter]{aa} 

\usepackage{graphicx}
\usepackage{txfonts}

\def\Msun{M$_{\sun}$}
\def\Mjup{M$_{\rm Jup}$}

\def\bp{$\beta$~Pic}
\def\au{{\sc au}}

\begin{document}

\title{A probable giant planet imaged in the $\beta$~Pictoris disk\thanks{Based on observations collected at the
European Southern Observatory, Chile, ESO; runs 072.C-0624(B) and
60.A-9026(A); and observations made with the Space Telescope Imaging
Spectrograph on board the NASA/ESA \emph{Hubble Space Telescope.}}}

\subtitle{VLT/NACO Deep $L$-band imaging}

\author{
  A.-M.~Lagrange \inst{1}
  \and
  D.~Gratadour \inst{2}
  \and
  G.~Chauvin \inst{1}
  \and
  T.~Fusco \inst{3}
  \and
  D.~Ehrenreich \inst{1}
  \and
  D.~Mouillet \inst{1}
  \and
  G.~Rousset \inst{2,3}
  \and
  D.~Rouan \inst{2}
  \and
  F.~Allard \inst{4}
  \and
  \'E.~Gendron \inst{2}
  \and
  J.~Charton \inst{1}
  \and
  L.~Mugnier \inst{3}
  \and
  P.~Rabou \inst{1}
  \and
  J.~Montri \inst{3}
  \and
  F.~Lacombe \inst{2}  
}

\offprints{A.-M. Lagrange}

\institute{
  Laboratoire d'Astrophysique de l'Observatoire de Grenoble,
  Universit\'e Joseph Fourier, CNRS, BP 53, 38041 Grenoble, France
  \email{anne-marie.lagrange@obs.ujf-grenoble.fr}
  \and
  LESIA, Observatoire de Paris, CNRS, 
  Universit\'e Pierre et Marie Curie, Universit\'e Denis Diderot,\\
  5 place Jules Janssen, 92190 Meudon, France
  \and
  Office National d'\'Etudes et de Recherches A\'erospatiales, 
  29 avenue de la Division Leclerc, 92322 Ch\^atillon, France
  \and
  Centre de Recherche Astronomique de Lyon, 
  46 all\'ee d'Italie, 69364 Lyon cedex 7, France
}

\date{Received date / Accepted date}

\abstract
{Since the discovery of its dusty disk in 1984,
\object{$\beta$~Pictoris} has become the prototype of young early-type
planetary systems, and there are now various indications that a
massive Jovian planet is orbiting the star at $\sim10$~\au. However,
no planets have been detected around this star so far.}
{Our goal was to investigate the close environement of \bp, searching
for planetary companion(s).}
{Deep adaptive-optics $L'$-band images of \bp\ were recorded using the
NaCo instrument at the Very Large Telescope.}
{A faint point-like signal is detected at a projected distance of
$\simeq 8$~\au\ from the star, within the North-East side of the dust
disk. Various tests were made to rule out with a good confidence level
possible instrumental or atmospheric artefacts. The probability of a
foreground or background contaminant is extremely low, based in
addition on the analysis of previous deep HST images. Its $L'=11.2$
apparent magnitude would indicate a typical temperature of $\sim
1500$~K and a mass of $\sim8$~\Mjup. If confirmed, it could explain
the main morphological and dynamical peculiarities of the \bp\
system. The present detection is unique among A-stars 
by the proximity of the resolved planet to its parent star. Its closeness and location
inside the \bp\ disk suggest a formation process by core accretion or
disk instabilities rather than binary like formation processes.}
{}


\maketitle

\section{Introduction}

Understanding planetary systems formation and evolution has become one
of the biggest challenges of astronomy, since the imaging of a debris
disk around $\beta$~Pictoris in the 80's (\cite{smith84}) and the
discovery of the first exoplanet around the solar-like star 51~Pegasi
during the 90's (\cite{mayor95}). While about 20 debris disks --
disks containing dust which is not primordial but produced by
collisions among larger rocky bodies -- have been resolved at optical
wavelengths today, $\beta$~Pic, a A5V star at a distance of $19.3\pm0.2$~pc
(\cite{crifo97}), remains the best studied young ($12^{+8}_{-4}$~Myr;
\cite{zuckerman01}) system, with an impressive amount of indirect
signs pointing toward the presence of planets.

The disk shows a relative inner void of matter inside
$50$~astronomical units (\au). Lecavelier des Etangs et al. (1995)
presented intriguing light variations possibly due to disk
inhomogeneities produced by a Jupiter size planet at $>6$~\au.  Several
asymmetries have been identified in the disk at optical
(\cite{kalas95}, \cite{heap00}) and infrared (\cite{telesco05})
wavelengths, as well as a warp at $\sim 50$~\au\ (\cite{mouillet97};
\cite{heap00}). The structure is well reproduced by the deformation
induced on colliding planetesimals by a giant planet on a slightly
inclined orbit within 50~\au\ from the star (\cite{krist96},
\cite{mouillet97}, \cite{gorka00}, \cite{augereau01} and
\cite{thebault01}). Silicate dust is observed as circumstellar rings
at 6, 16, and 30~\au\ from the star (\cite{okamoto04}), which could be
explained by the presence of a 2--5-Jovian-mass (\Mjup) planet at
$\sim 10$~\au\ from the star (\cite{freistetter07}). Evaporating star
grazing comets have been evidenced (see, e.g., \cite{lagrange00} for a
review) and dynamical simulations showed that the gravitational
perturbation of at least one giant planet at $\sim 10$~\au\ can
account for the observed rate of evaporating bodies
(\cite{beust00}). However, no planets have been detected so far,
either through direct imaging or through radial velocity studies, due
to the instrumental limitations of both techniques
(\cite{galland06}). In particular, the high spatial resolution imaging
detection capabilities were so far limited to distances typically $\ga
15$--$20$~\au.\\ We used the NAOS-CONICA instrument (NaCo), installed
on the Very Large Telescope UT4 (Yepun) set in Paranal (Chile), to
benefit from both the high image quality provided by the Nasmyth
Adaptive Optics System (NAOS; \cite{rousset03}) at infrared
wavelengths and the good dynamics offered by the Near-Infrared Imager
and Spectrograph (CONICA; \cite{lenzen03}) detector, in order to study
the immediate circumstellar environement of \bp.

\section{Observations and data reduction procedures}

\subsection{Observations}
\label{sec:obs} 

$L'$-band images of \bp\ ($V=3.8$, $L'=3.5$) were obtained between
2003 November 10 and 2003 November 17 with NaCo. The visible wavefront
sensor was used with the $14\times14$ lenslet array, together with the
visible dichroic.  We used the CONICA L27 camera, which provides a
pixel scale of $\sim 27$~mas.  Saturated images of \bp\ were recorded,
with detector integration times (DITs) of $0.175$~s and number of
detector integrations (NDIT) of $100$ or $200$. Every two
exposures\footnote{An exposure is completed after $\rm DIT \times
NDIT$.}, spatial offsets were applied in order to allow sky and
instrumental background removal. Non-saturated images were also
recorded to get images of the stellar point spread function (PSF) as
well as a photometric calibration. In such case, we added the Long
Neutral Density filter (transmission $\sim 0.018$) in the CONICA
optical path, and recorded images with DITs of $0.4$s. 

In addition, the binary IDS~22141S3712 (separation $\rho = 6\,630 \pm
10$~mas, position angle $\rm PA = 302.06 \pm 0.07\degr$;
\cite{vandessel93}) was observed on November~11 as an astrometric
calibrator. A mean plate scale of $27.105 \pm 0.041$~mas and a true
North orientation of $-0.10 \pm 0.07\degr$ were derived and used to
calibrate all \bp\ images. Saturated and non saturated exposures were
taken on the reference star \object{HR~2435} (A0II, $V=4.4$). The
purpose is to correct for the star halo (the wings of the PSF) present
in the saturated exposures. To optimize the removal of any fixed
speckle, \bp\ and HR~2435 were observed, as much as possible, at close
parallactic angles. Finally, twilight flat fields were also recorded
in $L'$ band.

The observing conditions varied from exceptional (coherent
energy\footnote{Since it is not possible to measure the Strehl ratio
on our saturated data, we can only use the information provided by
NAOS to assess the image quality. The coherent energies are those
measured by the system in $K$ band.} $EC > 70\%$, coherent time
$\tau_0 > 20$~ms) to reasonable ($EC \sim 50\%$, $\tau_0$ of a few
ms), and sometimes poor ($EC \sim 20$--$35\%$, $\tau_0 \sim
1$--$2$~ms), over the run; the data quality varies accordingly. The
best data set on \bp\ was obtained on November~10. In the following,
we will describe the data reduction and analysis of three
sets of data: (A) the very best set obtained on November 10; (B) a set
of data with slightly poorer image quality obtained on the same night,
with a shorter total exposure time, and (C) a set of data with poorer
image quality obtained on November 13. However, Set~C is
representative of the best data obtained in the nights following
November 10. The instrumental configurations and the status of image
quality for these three data sets can be found in Table~\ref{stats}.

\begin{table*}
\caption{Observing Log and companion position and flux relative to \bp\ for the 3 data set (A, B and C).}
\label{stats}
\centering
\renewcommand{\footnoterule}{}  

\begin{tabular}{l l l l l l l l l l l l l}
\hline \hline
\multicolumn{10}{c}{}  &   \multicolumn{3}{c}{\underline{\hspace{1.5cm}COMPANION\hspace{1.5cm}}}          \\
Set & Star & Date & DIT & NDIT & $t_\mathrm{exp} $ & $\pi$\footnote{Range of 
parallactic angles at the start/end of the observation.} & $\sec z$ & $\langle EC \rangle$
\footnote{The average coherent energy as estimated on-line by AO.} & $\langle\tau_0\rangle$\footnote{The average coherence time as estimated on-line by AO.} & Separation & PA & $\Delta L'$ \\
    &      &      &  (s)    &          & (s)                   & (o)             &     & (\%)      & (ms)        &   (mas)   & (o)   & (mag)  \\
\hline
A & \bp     & 2003-11-10 & 0.175 & 200 & 665 & $-18$/$-9$  & 1.125/1.119 & 62.71 & 11.47 & $411\pm8$ & $31.8\pm1.3$ & $7.7\pm0.3$\\
  & HR~2435 & 2003-11-10 & 0.175 & 200 & 630 & $-20$/$-12$ & 1.148/1.140 & 63.06 & 10.65 & & & \\
\hline
B & \bp     & 2003-11-10 & 0.175 & 100 & 420 & $-27$/$-19$ & 1.138/1.127 & 57.94 & 8.08 & $411\pm8$ & $31.5\pm1.3$ & $7.9\pm0.4$\\
  & HR~2435 & 2003-11-10 & 0.175 & 100 & 420 & $-28$/$-20$ & 1.161/1.149 & 60.23 & 9.70 & & & \\
\hline
C & \bp     & 2003-11-13 & 0.175 & 200 & 385 & $-24$/$-19$ & 1.161/1.149 & 46.94 & 4.50 & $401\pm8$ & $32.1\pm1.4$ & $7.6\pm0.4$ \\
  & HR~2435 & 2003-11-13 & 0.175 & 200 & 350 & $-22$/$-17$ & 1.150/1.144 & 44.99 & 2.72 & & & \\
\hline
\end{tabular}
\begin{list}{}{}
\item[$^3$] Range of parallactic angles at the start/end of the observation. 
\item[$^4$] The average coherent energy as estimated on-line by AO. 
\item[$^5$] The average coherence time as estimated on-line by AO. 
\end{list}
\end{table*}

\subsection{Data processing}

The first step was the cosmetic correction (bad-pixels, flat-fielding,
background subtraction) and recentering of individual offset positions
of \bp\ and HR\,2435 observations. A first method was to directly
apply standard routines from the \texttt{eclipse} library
(\cite{devillard97}), using classical cross-correlation algorithm. A
special care was taken to estimate the background at each given
position by averaging images obtained at the previous and successive
offset position in the observing sequence. Alternatively, a second
method was used with an improved software dedicated to adaptive optics
(AO) image processing (see \cite{gratadour05}). Following a different
approach for the background subtraction, images at individual offset
positions were recentered using a maximum likelihood algorithm at the
level of the tenth of a pixel or better. The same overall process was
used for both the object and the reference. Consistent results are
found in terms of recentering precision and background subtraction (at
less than the backgroung noise of 0.9~ADU).

\begin{figure}[t]
  \centering
  \includegraphics[width=\hsize]{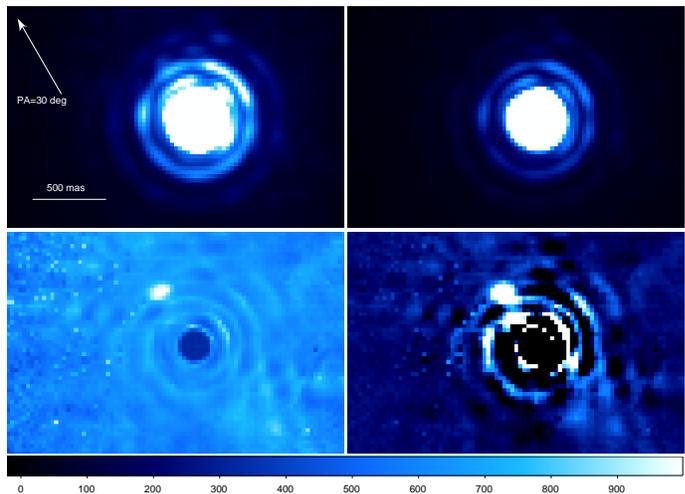}
  \caption{\bp\ and HR\,2435 recentered and saturated $L'$ images (top
  left and top right, respectively) in data set A. Below are the
  divided (bottom left) and subtracted (bottom right) images. North is
  up and East is to the left. A candidate companion is clearly
  detected at a PA of $\simeq 32\degr$, i.e., along the NE side of the
  disk, at a separation of about 0\farcs41 from the star.}
  \label{sous_ndit200_nov10}
\end{figure}

As a second step, three parallel approaches were followed to study 
the close environement of \bp:

\begin{itemize}

\item first approach consists in removing the PSF wings from the
saturated images of \bp\ by a simple minimization of the residuals.
To do so, we first divided \bp\ images by the ones of HR~2435 obtained
under similar conditions (same DIT, NDIT, and pupil position). We then
computed the scaling factor to be applied to the reference in order to
scale its flux to that of \bp. We then subtracted the reference
images to the \bp\ ones. Recentering and scaling processes are
repeated to minimize the residuals with respective precisions
of sub-pixels and 5\%. Tests have been performed
using rotated images of \bp\ for PSF subtraction. However, the fact
that PSF halo is not centro-symmetric and that the statical
aberrations are not overlaid worsen the subtraction process.

\item the second approach is to follow the same subtraction sequence
but applying the maximum likelihood algorithm of \cite{gratadour05}
(see Fig.~\ref{sous_ndit200_nov10}). The algorithm is used here to
recenter the reference star with \bp\ and therefore confirm the
previous estimation. Similar precisions are achieved.

\item the last approach was actually to use the MISTRAL deconvolution
algorithm (\cite{mugnier04}), based on a maximum a posteriori
scheme. Nevertheless, \texttt{MISTRAL} relies on a strict convolution
process between image and reference which is not the case for our
saturated data. A first step is therefore to perform a posteriori
correction of saturated part of the image and reference. This is done
using a simulated Airy pattern. The top of the Airy pattern replaces
the image saturated pixels. The flux level is adjusted using the
first Airy rings. Such a correction is possible because of the very
good Strehl ratio on the image. Meanwhile, if this a posteriori
correction does not significantly affect the restitution of the object
structures, it could obviously degrade the relative photometry. Hence,
the deconvolution process is an alternative approach (compared to
reference subtraction and division) for the image processing which is
less sensitive to image versus reference centering. It allows us to
provide the best measurements of relative position with a precision of
0.3 pixels, but remain uncertain for the relative photometry due to
the use of saturated images.
\end{itemize}

\section{A candidate companion in the \bp\ disk?}

\subsection{Results from the highest-quality data (set A)}

Using three independent approaches, the companion detection is
confirmed at the same location. The resulting images, using the
maximum likelihood algorithm for recentering, are reported in
Figure~\ref{sous_ndit200_nov10}.  The  companion candidate (hereafter,
the CC) point-like signal is clearly visible in the divided and subtracted images.  The maximum of the signal is about $190$~ADU. Different techniques
(variable aperture photometry, 2D-gaussian fitting, PSF-fitting) were
used to extract the CC flux, giving consistent results. As the
reference recentering and rescaling actually dominate the flux
measurement precision, flux uncertainties were derived considering
respective variations of 0.3 pixel and 5\% in the subtraction
process. We obtain a contrast of $\Delta L' = 7.7\pm0.3$ between the
CC and \bp. Using deconvolution, we derive a separation of
$411\pm8$~mas and a PA of $31.8 \pm 1.3\degr$ relative to the primary,
i.e along the NE side of the disk.
\begin{figure}
  \centering
  \includegraphics[width=\hsize]{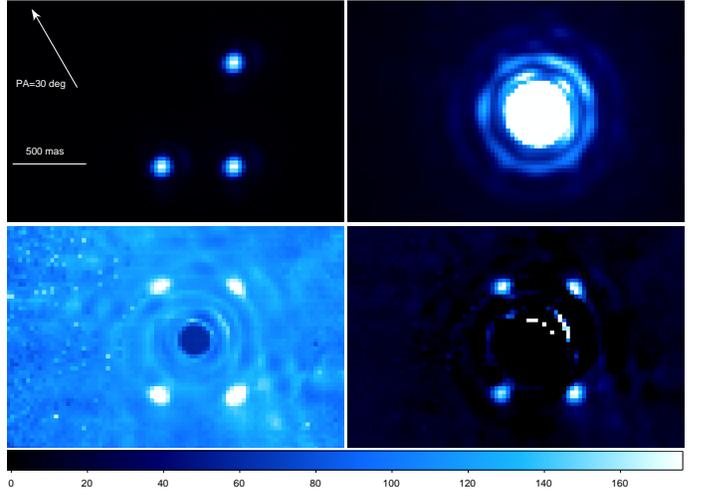}
  \caption{Top left: simulated planets at PA of $150\degr$, $210\degr$
  and $330\degr$. Top right: composite image of \bp\ plus the fake
  planets. Bottom left: division of the composite image by the
  saturated image of HR~2435. Bottom right: scaled subtraction of the
  composite image by the saturated image of HR~2435. Note that even a
  slight (0.3 pixel) relative offset between \bp\ and HR~2435 impacts
  the resulting shape of the fake planets as much as the candidate
  one. In particular, triangular shapes can be observed, due to the
  proximity of the slightly inner Airy ring.}
  \label{test_planet_set1}
\end{figure}

The use of different methods excludes artefacts created during the
reduction process. In particular, the result of the deconvolution
rules out any effect that could be introduced by imperfect estimation
of the offset between \bp\ and HR~2435 saturated images due, for
instance, to a possible contribution of the disk. We did check anyway
that the disk signal is very faint and not significantly
asymmetric. To rule out detector effects, we looked for possible
remanence and electronic ghosts that could occur because our images
are saturated. Individual images inspection exclude any contamination
by these two effects that rapidly disappear after a few
frames. Artefacts due to the very good but still imperfect AO
correction are still possible. However, aberrations due to a
modulation of the deformable mirror would generally lead to the
presence of either symmetrical or anti-symmetrical patterns. Static
aberrations should be present equivalently around \bp\ and HR~2435, as
both stars were observed with similar pupil configurations. We have
then tested the possible impact of an imperfect removal of static
speckles due to the variation of the parallactic angle during the
observations of \bp\ and HR~2435 (up to $8\degr$). We processed
individual pairs of data of \bp\ and HR~2435 taken with parallactic
angles equal within $\pm 0.4\degr$ and added up the individual
subtracted images. The CC is still present and appears slightly
sharper (but still compatible with the instrumental resolution). In
addition, since the same signal is also observed a few nights apart
(see below), we conclude that quasi-static aberrations are unlikely.

To further assess the reality of the detection and test the CC
photometry, we added three `fake planets' at similar separations but
different PA ($150\degr$, $210\degr$, and $350\degr$) to the
recentered and stacked image of \bp. The fake planet images were
generated by scaling and shifting an unsaturated image of \bp\ taken
during the same night. To match the level of the observed signal, the
magnitudes of the fake planets are scaled to the measured flux ratio
on set A. We then subtracted the scaled image of HR~2435 to that of
the composite image. The result is shown in
Figure~\ref{test_planet_set1}, from which it is clear that the fake
planets produce features similar to the observed signal, supporting
our contrast estimate.  We therefore, derive an apparente magnitude of
$L'=11.2\pm0.3$ for the CC.

\subsection{Results from lower quality data (sets B and C)}

\begin{figure}
  \centering
  \includegraphics[width=\hsize]{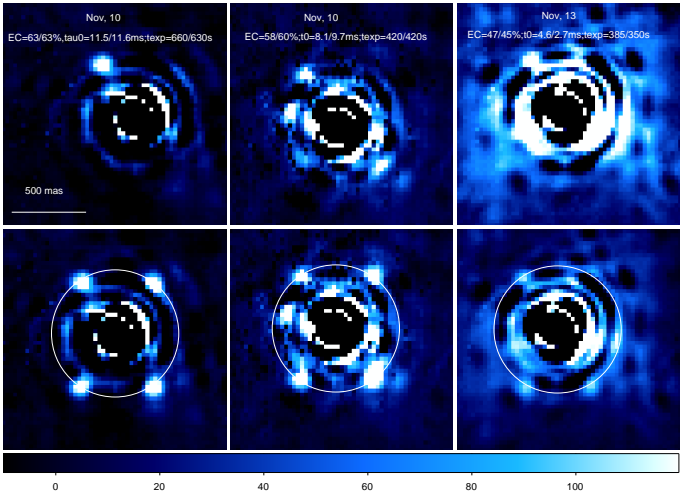}
  \caption{Top: Scaled subtraction of the composite image by the
    saturated image of HR~2435 for sets A, B and C. Bottom: Simulated
    planets at PA of $150\degr$, $210\degr$ and $330\degr$.Note that even
    a slight (0.3 pixel) relative offset between \bp\ and HR~2435 impacts
    the resulting shape of the fake planets as much as the candidate
    one. In particular, triangular shapes can be observed, due to the
    proximity of the slightly inner Airy ring. }
  \label{sous_allsets_plussimus}
\end{figure}

All other sets of data were processed as before. It quickly appeared
that the detection becomes more marginal when the exposure time and/or
the image quality decrease. For the best remaining sets in terms of
atmospheric conditions and relative parallactic angle between \bp\ and
HR~2435 (described in Section~\ref{sec:obs} and Table~\ref{stats}),
the results are shown in Figure~\ref{sous_allsets_plussimus}. The same
point-like signal is seen in both cases, with however a lower
signal-to-noise ratio. This is due to the lower exposure times and the
(slightly) poorer image qualities. This is illustrated in
 Figure~\ref{test_planet_set1}, where we also added fake planets 
with a magnitudes equal to that of the fake planets used for Set A 
 to the data sets B and C. The fake planet signals are, as the CC
signal, much less detectable under these conditions; hence the results
from our 3 data sets are consistent.

\section{Bound companion or background object?}

With these data alone, it is not excluded that the CC could be a
foreground or background object. To definitely test this hypothesis,
second epoch measurements are needed. Using the known \bp\
proper and parallactic motions, we have considered the position of the
CC assuming it would be a stationary contaminant. It would be closer
than $0\farcs2$ to \bp\ in 2008 and hence, not detectable with current
technics. In past years, $\beta$~Pic has been monitored by numerous
programs of the \emph{Hubble Space Telescope}, including observations
using the Advanced Camera for Surveys (Golimowski et al.\ 2006 and
Kalas et al.\ 2005), the Near-Infrared Camera and Multi-Object
Spectrometer (Brown et al.\ 1999) and the Wide Field Planetary Cameras
(Kalas et al.\ 2000). The lack of spatial resolution or the size of
the coronographic mask prevented from getting any reliable hints
of any point sources close to the star.

To our knowledge, the highest angular resolution and dynamical data
available are the coronagraphic data taken in 1997 September with STIS
(\cite{heap00}). These authors were able to probe the close stellar
environment down to 0\farcs75 from \bp. They did not report any CC in
their images at the expected location of 0\farcs9 North and 0\farcs25
East to $\beta$~Pic. Based on STIS detection limits and the brightness
derived by Heap et al. (2000), all objects brighter $V \leq 17$ would
have been detected. Red giants or supergiants can be ruled out for
their distances that would bring them unrealistically far (typically $\ga 10\,000$~pc) from us. If
we consider a foreground or background stellar field contaminant, it
would therefore have a large ($V-L\geq5.8$) color, i.e a spectral type
later than mid-M. Based on a number density of L dwarfs of
$1.9\times10^{-3}$~pc$^{-3}$ given by Burgasser (2001) and Cruz et
al. (2003), the probability of finding a foreground or background
field L dwarf in a region of 500~mas radius around \bp\ and with
$L'=11.2$ is about~$10^{-10}$. Without any assumption on the
contaminant spectral type, one can use galactic population model
outputs to estimate the probability to find any ($L'\leq12$) galactic
source at 500~mas from \bp. We find a low probability of $6
\times10^{-5}$. Last, we cannot strictly rule out a contamination by
an extragalactic source, such as a high-redshift quasar. In
conclusion, a contamination appears very unlikely. In addition, the
fact that the candidate companion falls into the disk strongly favors
that it is bound to the star.

\section{Implications on the understanding of the \bp\ system}

With an $L'$ magnitude of $11.2\pm0.3$, and assuming a
distance of $19.3\pm0.2$~pc and an age of $12^{+8}_{-4}$~Myr
(\cite{zuckerman01}), the mass of the CC is estimated to
$9^{+3}_{-2}$~\Mjup\ using COND models (\cite{baraffe03}) and
$8^{+4}_{-2}$~\Mjup\ using DUSTY models (\cite{chabrier00}). The
planet should still be in the phase of cooling: Dusty and COND models
predict effective temperatures of 1\,400 and 1\,600~K,
respectively. The validity of these models in the case of young
planets formed by core accretion has been recently questionned
(\cite{marley08}) on the basis that the accretion shock might impact
the planet initial internal entropy and its subsequent early thermal
evolution. These authors claim that young giant planets could be
significantly cooler and fainter than predicted so far. 
However, the treatment of the accretion shock is still a matter of debate. The
impact of the initial disk mass is also to be studied. This may be
important in the present case as \bp\ is significantly more massive
than the Sun. In the present case,  \cite{marley08} model predicts a luminosity ten times fainter from what we observed for an $8$~\Mjup\ planet at the age of \bp. However, we can already note that a companion at a true
separation of $\sim 8$~\au\ could not be significantly more massive
than 10--20~\Mjup\ since otherwise it would have been detected through
radial velocity measurements, as shown by a detailed analysis that
will be presented in a forthcoming paper (Desort et al., in
preparation).

If the observed projected separation happens to be the physical one or
close to the physical one, then the companion explains three very
intriguing and so far unique characteristics of the \bp\ system: the
warp in the inner disk, the inner belts and the falling evaporating
bodies (\cite{freistetter07}). Mouillet et al.\ (1997) and Heap et
al.\ (2000) showed that the warp of the disk constrains the (mass
$M_p$, semi-major axis $a$) domain of the planet as follows:
\begin{equation}
\log (M_p / M_\star) + 2 \log a + \log t \approx 6.7,
\end{equation}
where $M_\star = 1.8$~\Msun\ (\cite{crifo97}) and $t$ is the age of
the star. Given an age $t = 12$~Myr, masses $M_p = 6$ and $13$~\Mjup\
give $a = 9.7$ and $7.6$~\au, respectively, nicely bracketting the
measured projected separation. This further strengthens the likelyhood
that the observed CC \emph{is} the planet causing the warp.

At the time of the submission of this letter, this companion was to
our knowledge the first one possibly detected around an A type MS
star.  Recent results published since then by Marois et al. (2008) and
Kalas et al. (2008) show also the presence of planets at distances 24
to 118 AUs from two A types stars. This companion could be anyway the
first extrasolar planet ever imaged so close to its parent star:
8~\au. In particular, it would be located well inside the orbits of
the outer planets of the Solar System. Its closeness and location
inside the \bp\ disk suggest a formation process by core accretion or
disk instabilities rather than binary like formation mechanisms
proposed for the companions to 2M1207 and AB\,Pic (Chauvin et
al. 2005a, b). Further direct imaging observations should allow to
constrain the planet orbit and hence measure the planet dynamical
mass. This would in turn allow to test the models of planet
formation. This is dramatically needed as the evolutionary models are
not calibrated with real data in these ranges of masses and
ages. Finally, we note that unfortunately, if the projected separation
is the physical one, the planet orbital period is about 16~years and
should be at present time much closer to the star and hence
undetectable with NaCo.

\begin{acknowledgements}

We acknowledge financial support from the French Programme National de
Plan\'etologie (PNP, INSU), as well as from the French Agence
Nationale pour la Recherche (ANR; project $NT05-4_44463$). These results have made use of the
SIMBAD database, operated at CDS, Strasbourg, France. We also thank
the ESO staff for its help during the NaCo observations. We thank
J.-C.~Augereau for useful discussions on the \bp disk.  
AML thanks also P.~Rubini for his
help on the lay out of the paper. Finally, we thank the referee for
his rapid and precious comments on the paper.
\end{acknowledgements}


\begin{thebibliography}{}
\bibitem[Augereau et al.\ 2001] {augereau01} Augereau, J.-C., Nelson, R.~P., Lagrange, A.-M., Papaloizou, J.~C.~B., \& Mouillet, D.\ 2001, A\&A, 370, 447
\bibitem[Baraffe et al.\ 2003]{baraffe03} Baraffe, I., Chabrier, G., Barman, T.~S., Allard, F., \& Hauschildt, P.~H.\ 2003, A\&A, 402, 701
\bibitem[Beust \& Morbidelli 2000] {beust00} Beust, H., \& Morbidelli, A.\ 2000, Icarus 143, 170
\bibitem[Brown et al.\ 1999]{brown99} Brown, J., Bennum, D., Schultz, A.~B., et al.\ 1999, Bulletin of the AAS, 31, 1487
\bibitem[Burgasser 2001]{burgasser01} Burgasser, A.\ 2001, PhD Thesis, California Institute of Technology
\bibitem[Chabrier et al.\ 2000] {chabrier00} Chabrier, G., Baraffe, I., Allard, F., \& Hauschildt, P.~H.\ 2000, ApJ, 542, 464
\bibitem[Chauvin et al.\ 2005a]{chauvin05a} Chauvin G., Lagrange A.-M., Dumas C. et al,, 2005, A\&A, 438, L25
\bibitem[Chauvin et al.\ 2005b]{chauvin05b} Chauvin G., Lagrange A.-M., Zuckerman B. et al. 2005, A\&A, 438, L29
\bibitem[Crifo et al.\ 1997] {crifo97} Crifo, F., Vidal-Madjar, A., Lallement, R., Ferlet, R., \& Gerbaldi, M.\ 1997, A\&A, 320, L29
\bibitem[Cruz et al.\ 2003]{cruz03} Cruz, K.~L., Reid, I.~N., Liebert, J., Kirkpatrick, J.~D., \& Lowrance, P.~J.\ 2003, AJ, 126, 2421
\bibitem[Freistetter et al.\ 2007] {freistetter07} Freistetter, F., Krivov, A.~V., \& L\"ohne, T.\ 2007, A\&A, 466, 389
\bibitem[Devillard 1997]{devillard97} Devillard N. 1997, Messenger, 87
\bibitem[Ducati et al.\ 2001] {ducati01} Ducati, J.~R., Bevilacqua, C.~M., Rembold, S.~B., \& Ribeiro, D.\ 2001, ApJ, 558, 309
\bibitem[Galland et al.\ 2006] {galland06} Galland, F., Lagrange, A.-M., Udry, S., et al.\ 2006, A\&A, 447, 355
\bibitem[Golimowski et al.\ 2006]{golimowski06} Golimowski, D.~A., Ardila, D.~R., Krist, J.~E., et al.\ 2006, AJ, 131, 3109
\bibitem[Gorkavyi et al.\ 2000] {gorka00} Gorkavyi, N., Heap, S., Ozernoy, L., Taidakova, T., Mather, J. 2000, astro.ph.12470
\bibitem[Gratadour et al.\ 2005] {gratadour05} Gratadour, D., Mugnier, L.~M., \& Rouan, D.\ 2005, A\&A, 443, 357
\bibitem[Heap et al.\ 2000] {heap00} Heap, S.~R., Lindler, D.~J., Lanz, T.~M., et al.\ 2000, ApJ, 539, 435
\bibitem[Lecavelier des Etangs et al.\ 1995] {lecavelier95} Lecavelier des Etangs, A., Deleuil, M., Vidal-Madjar, A., et al.\ 1995, A\&A, 299, 557
\bibitem[Kalas \& Jewitt 1995] {kalas95} Kalas, P., \& Jewitt, D.\ 1995, AJ, 110, 794
\bibitem[Kalas et al.\ 2000]{kalas00} Kalas, P., Larwood, J., Smith, B.~A., \& Schultz, A.\ 2000, ApJ, 530, L133
\bibitem[Kalas et al.\ 2005]{kalas05} Kalas, P., Graham, J.~R., \& Clampin, M.\ 2005, Nature, 435, 1067 
\bibitem[Kalas et al.\ 2008] {kalas08} Kalas, P, Graham, J.R., Chiang, E., Fitzgerald, M.P., Clampin, M., Kite, E.S., Stapelfeldt, K., Marois, C., Krist, J. 2008, arXiv0811.1994K
\bibitem[Krist et al.\ 1996] {krist96} Krist, J.E., Burrows, C.J., Stapelfeldt, K.R., Watson, A.M.,\& WFPC2 Investigation Definition Team 1995, AAS, 187, 4413
\bibitem[Lafreni\`ere et al.\ 2008]{lafreniere08} Lafreni\`ere, D., Doyon, R., Marois, C., et al.\ 2008, ApJ, 670, 1367
\bibitem[Lagrange et al.\ 2000] {lagrange00} Lagrange, A.-M., Backman, D.~E., \& Artymowicz, P.\ 2000, Planetary Material around Main-Sequence Stars, in Protostars and Planets {\sc iv}, ed.\ V.~Mannings, A.~P.~Boss, \& S.~S.~Russell (University of Arizona Press, Tucson), 639
\bibitem[Leggett et al.\ 2002]{leggett02} Leggett, S.~K., Golimowski, D.~A., Fan, X., et al.\ 2002, ApJ, 564, 452
\bibitem[Lenzen et al.\ 2003] {lenzen03} Lenzen, R., Hartung, M., Brandner, W., et al.\ 2003, SPIE 4841, 944
\bibitem[Luhman et al.\ 2006]{luhman06} Luhman, K.~L., Wilson, J.~C., Brandner, W., et al.\ 2006, ApJ, 649, 894
\bibitem[Metchev \& Hillenbrand 2006]{metchev06} Metchev, S., \& Hillenbrand, L.\ 2006, ApJ, 651, 1166
\bibitem[Mugnier et al.\ 2004] {mugnier04} Mugnier, L.~M., Fusco, T., \& Conan, J.-M.\ 2004, JOSAA, 21, 1841
\bibitem[Smith \& Terrile 1984] {smith84} Smith, B., \& Terrile, R.\ 1984, Science, 226, 1421
\bibitem[Marley et al.\ 2007] {marley08} Marley, M.~S., Fortney, J.~J., Hubickyj, O., Bodenheimer, P., \& Lissauer, J.~J.\ 2007, ApJ, 655, 541
\bibitem[Marois et al.\ 2007] {marois08} MArois, C., MacIntosh, B., Barman, T., Zuckerman, B., Inseok, S., Patience, J., LAfreniere, D., Doyon, R. 2008, Science-express
\bibitem[Mayor \& Queloz 1995] {mayor95} Mayor, M., \& Queloz, D.\ 1995, Nature, 378, 355
\bibitem[Mouillet et al.\ 1997] {mouillet97} Mouillet, D., Larwood, J.~D., Papaloizou, J.~C.~B., \& Lagrange, A.-M.\ 1997, MNRAS, 292, 896
\bibitem[Okamoto et al.\ 2004] {okamoto04} Okamoto, Y.~K., Kataza, H., Honda, M., et al.\ 2004, Nature, 431, 660
\bibitem[Rousset et al.\ 2003]{rousset03} Rousset, G., Lacombe, F., Puget, P., et al.\ 2003, SPIE 4839, 140
\bibitem[Telesco et al.\ 2005] {telesco05} Telesco, C.~M., Fisher, R.~S., Wyatt, M.~C., et al.\ 2005, Nature, 433, 133
\bibitem[Th\'ebault \& Beust 2001] {thebault01} Thébault, P. \& Beust, H. 2001, A\&A, 376, 621
\bibitem[van~Dessel \& Sinachopoulos 1993]{vandessel93} van Dessel, E., \& Sinachopoulos, D.\ 1993, A\&AS, 100, 517
\bibitem[Woodgate et al.\ 1998] {woodgate98} Woodgate, B.~E., Kimble, R.~A., Bowers, C.~W., et al.\ 1998, PASP, 110, 1183
\bibitem[Zuckerman et al.\ 2001] {zuckerman01} Zuckerman, B., Song, I., Bessel, M.~S., \& Webb, R.~A.\ 2001, ApJ, 562, L87
\end{thebibliography}
\end{document}